\documentclass[aps, pre, twocolumn, superscriptaddress, amsmath,amssymb,floatfix]{revtex4}

\newif\ifpdf\ifx\pdfoutput\undefined\pdffalse\else\pdfoutput=1\pdftrue\fi

\usepackage{graphicx}
\usepackage{bm}
\usepackage{epsfig}

\newcommand{\be}{\begin{equation}}

\newcommand{\ee}{\end{equation}}

\begin{document}

\title{\bf Influence of polydispersity on the critical parameters of
an effective potential model for asymmetric hard sphere mixtures}

\author{Julio Largo}
\author{Nigel B. Wilding}
\affiliation{Department of Physics, University of Bath, Bath BA2 7AY, United Kingdom}

\date{\today}

\begin{abstract} 

We report a Monte Carlo simulation study of the properties of highly
asymmetric binary hard sphere mixtures. This system is treated within
an effective fluid approximation in which the large particles interact
through a depletion potential (R. Roth {\em et al}, Phys. Rev. E{\bf 62}
5360 (2000)) designed to capture the effects of a virtual sea of small
particles. We generalize this depletion potential to include the effects
of explicit size dispersity in the large particles and consider the case
in which the particle diameters are distributed according to a Schulz
form having degree of polydispersity $14\%$. The resulting alteration
(with respect to the monodisperse limit) of the metastable fluid-fluid
critical point parameters is determined for two values of the ratio of
the diameters of the small and large particles:
$q\equiv\sigma_s/\bar\sigma_b=0.1$ and $q=0.05$. We find that inclusion
of polydispersity moves the critical point to lower reservoir volume
fractions of the small particles and high volume fractions of the large
ones. The estimated critical point parameters are found to be in good
agreement with those predicted by a generalized corresponding states
argument which provides a link to the known critical adhesion parameter
of the adhesive hard sphere model. Finite-size scaling estimates of the
cluster percolation line in the one phase fluid region indicate that
inclusion of polydispersity moves the critical point deeper into the
percolating regime. This suggests that phase separation is more likely
to be preempted by dynamical arrest in polydisperse systems. 


\end{abstract} 
\maketitle
\setcounter{totalnumber}{10}

\section{Introduction and background}

\label{sec:intro}

Highly asymmetric mixtures of hard spheres have long served as a
prototype model for systems of large colloidal particles dispersed in a
sea of smaller colloids. A key physical feature of such systems is the
mediation by the small particles of so-called depletion forces between
the large ones \cite{BELLONI00}. This force has its origin in entropic
effects associated with the dependence of the free volume of the small
particles on the degree of clustering of the large ones. Although, the
typical range of the forces is rather limited (of order the diameter of
the small particles), they can be very strong. 

A longstanding issue in this context concerns the ability of depletion
forces to engender phase transitions in binary hard sphere mixtures.
Biben and Hansen \cite{BIBEN91} addressed this matter using integral
equation theory, and predicted that for sufficient size asymmetry,
depletion forces engender a fluid-fluid spinodal instability. Other
theoretical studies have arrived at often conflicting conclusions in
this regard (see discussion in ref.~\cite{DIJKSTRA99}), but experiments
on colloidal systems (eg. ref.~\cite{POON02}), do apparently confirm a
transition, although in certain circumstances it is found to be
metastable with respect to a broad fluid-solid coexistence region. 

Ideally one would like to settle the matter of the existence of a
fluid-fluid transition (as well as its stability or otherwise with
respect to the fluid-solid boundary), by computer simulation.
Unfortunately, direct simulation studies of very asymmetric additive
mixtures are severely hampered by extremely slow relaxation.
Accordingly, all such studies to date have been restricted to mixtures
of relatively low asymmetries, for which a fluid-fluid transition is
less likely to be observable. While recently developed novel algorithms
\cite{BUHOT,LUIJTEN05} offer some hope of future progress in accessing
greater size asymmetries, no direct evidence has (to date) been
obtained for the existence of a fluid-fluid phase separation in
additive hard sphere mixtures.

In view of these difficulties, a fruitful alternative to simulations of
the full two component mixture is to attempt to map it onto an effective
one-component system which can be simulated more easily. This
``effective fluid'' approach was taken by Dijkstra, van Roij and Evans
\cite{DIJKSTRA99}, and by Almarza amd Enciso \cite{ALMARZA99} who
proposed a model depletion potential by tracing out from the partition
function the degrees of freedom associated with the small particles. The
resulting interparticle potential for the large particles is
parameterized by the size ratio $q=\sigma_s/\sigma_b$ between small and
large particles, and a {\em reservoir} volume fraction of the small
particles $\tilde\eta_s$. Simulation-based free energy measurement of
the resulting system yielded, for values of $q<0.1$, a fluid-fluid
separation that was metastable with respect to a broad solid-fluid
coexistence region. Explicit simulations of the two component mixture
confirmed the accuracy of the model depletion potential as far as the
location of solid-fluid and solid-solid transitions was concerned, but
could not access the likely region of fluid-fluid separation. Subsequent
work by Roth, Evans and Dietrich \cite{ROTH00} yielded a more accurate
depletion potential by fitting to accurate DFT predictions. However to
our knowledge the latter potential has to date not been used to study
phase behaviour.

Real colloidal fluids are polydisperse, that is their constituent
particles exhibit an essentially continuous range of size, shape or
charge. Introducing polydispersity into model fluids is known to alter
fluid-fluid critical point parameters \cite{BAUS00,WILDING04A} as well
as freezing boundaries \cite{FASOLO04,KOFKE99}. It is therefore
pertinent to enquire as to the effect of polydispersity on the location
of the metastable fluid-fluid transition of hard sphere mixtures. Indeed
this question has previously been addressed in part by Warren
\cite{WARREN99} who applied the moment free energy method
\cite{SOLLICH98} to study a polydisperse version of the equation of
state of Boublik and Mansoori \cite{BOUBLIK} for binary hard sphere
mixtures. The results showed that for sufficiently large size ratio and
polydispersity of the large particles, a fluid-fluid spinodal appears in
the model. The transition was predicted to become more stable with
increasing degree of polydispersity. 

Most other theoretical investigations of polydispersity in hard sphere
mixtures \cite{MAO95B,GOULDING01,FRYDEL05} have focussed on the form of
the depletion potential and did not explicitly consider the
consequences for phase equilibria. The sole study of phase behaviour to
date (of which we are aware) is that of Fasolo and Sollich
\cite{FASOLO04} who applied the moment free energy method to the
Asakura-Oosawa (AO) model \cite{ASAKURA54}. This model describes the
limit of maximum non-additivity of the small particles and in contrast
to additive mixtures, the monodisperse AO model is known to exhibit a
{\em stable} fluid-fluid phase transition for size ratios $q\lesssim
0.5$ \cite{DIJKSTRA99B,VINK04}. The introduction of size polydispersity
to the large spheres \cite{FASOLO04} was observed to disfavor both
fluid-fluid and fluid-solid phase separation, though the effect was
larger for the latter transition. The net result was a lowering of the
$q$ value necessary for occurrence of stable fluid-fluid phase
separation, and an increase in the stability of this transition with
respect to freezing.

In the present work we apply specialized Monte Carlo simulation
techniques to an effective fluid model for additive mixtures with a
view to elucidating the effect of large particle (colloidal)
polydispersity on the parameters of the fluid-fluid critical point. Our
results indicate that polydispersity shifts the critical point to lower
reservoir volume fractions $\tilde\eta_s$ of the smaller particles, and
to higher volume fractions $\eta_b$ of the large ones. A determination
of the percolation threshold using finite-size scaling methods shows
that the addition of polydispersity moves the fluid-fluid critical
point deeper into the percolation region. We further find that accurate
predictions for the critical reservior volume fraction
$\tilde\eta_s^{\rm crit}$ can be obtained by matching the second virial
coefficient of the depletion potential to that of the adhesive hard
sphere model at its (independently known) critical point. 

 \section{Models}

\subsection{Depletion potentials}
\label{sec:pots}

Two model depletion potentials are considered in this work. The
first, on which we shall focus primarily, is due to Roth, Evans and
Dietrich \cite{ROTH00}, and we shall refer to it as the RED potential.
It derives from accurate density functional theory (DFT) studies of
hard sphere mixtures and takes the form:

\begin{equation}
W=\epsilon\frac{R_b+R_s}{2R_s} \bar{W}\:.
\label{eq:rothpot}
\end{equation}
Here $R_b$ and $R_s$ are the radii of the large and small spheres respectively. The
scaled depletion potential ${\bar W}={\bar W}(x,\eta_s)$ is a function
of $x=h/\sigma_s$, the distance from contact measured in units of the
small sphere diameter, and the volume fraction $\eta_s$
of the small spheres.. The parameter $\epsilon$ takes the value
$\epsilon=2$ for wall-sphere interactions and $\epsilon=1$ for
sphere-sphere interactions.

\begin{figure}[h]
\includegraphics[type=eps,ext=.eps,read=.eps,width=8.0cm,clip=true]{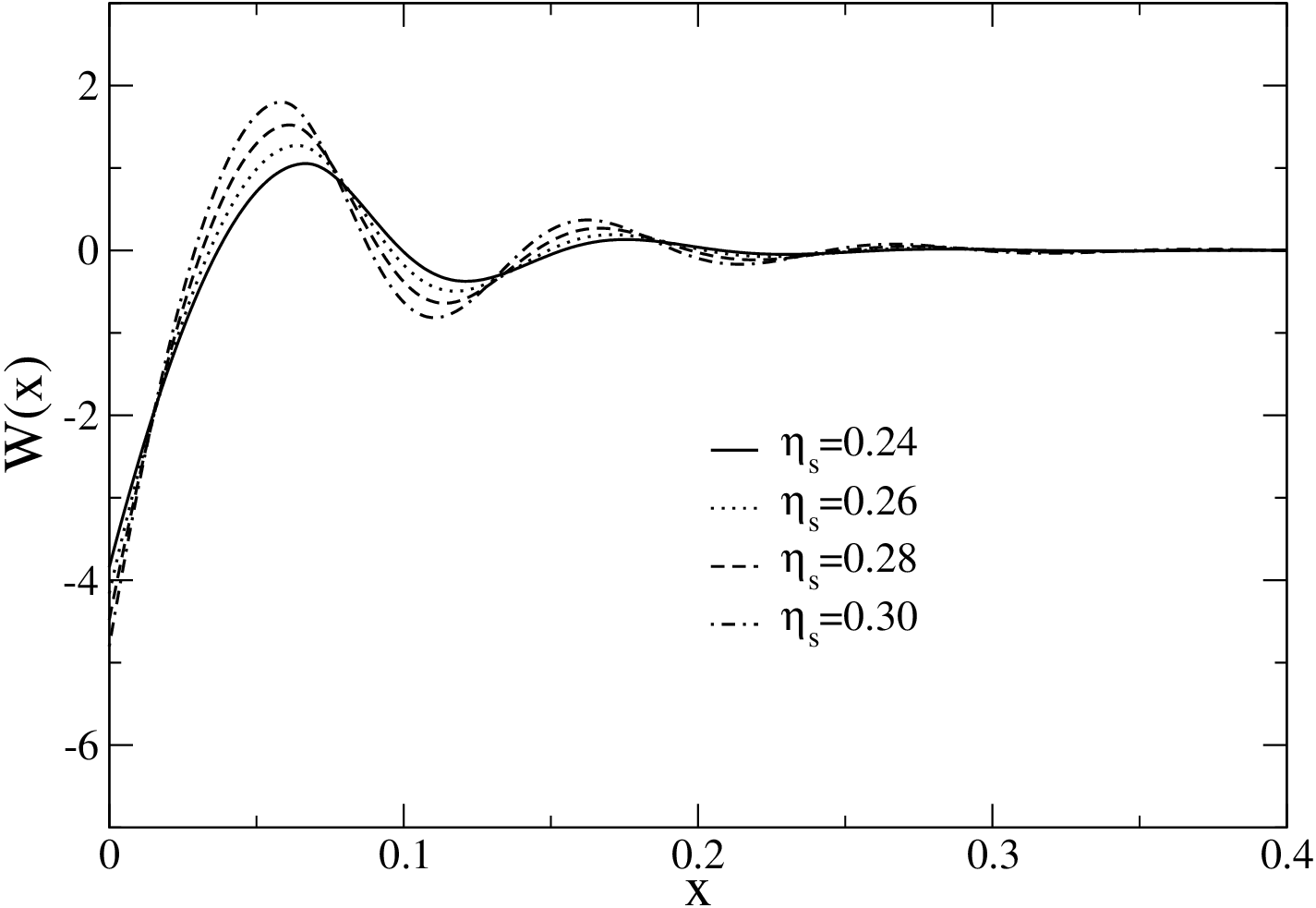}
\caption{The form of the RED potential for $q=0.1$ at a selection of values of $\eta_s$.} 
\label{fig:rothpot}
\end{figure}

Between contact at $x=0$ and the location of the first maximum $x_0$
the scaled depletion potential is expressed in terms of a cubic polynomial:

\begin{equation}
\beta \bar{W}(x, \eta_s)= a(\eta_s)+b(\eta_s) x+c(\eta_s) x^2 +d(\eta_s) x^3, \hspace{2mm} x \leq x_0\:.
\end{equation}
The coefficients $a,b,c$ and $d$ were obtained by Roth {\em et
al} by fitting to depletion potentials calculated within DFT. For
$x>x_0$, they assume that the asymptotic
regime already sets in. For the interaction between two spheres (a
different expression applies to the asymptotic behaviour between a sphere and a
hard wall) this is
 
\begin{equation}
\beta \bar{W}(x, \eta_s)= \beta \bar{W}^{\rm asym}(x,\eta_s), \hspace{2mm}x>x_0
\end{equation}
where 

\begin{eqnarray}
\label{eq:asympt}
\beta \bar{W}^{\rm asym}(x,\eta_s)&=&\frac{A_p(\eta_s)}{\sigma_s^{-1}(\sigma_b+h)}\exp(-a_0(\eta_s) \sigma_s x) \\ \nonumber
&\;&\times  \cos[a_1(\eta_s) \sigma_s x- \Theta_p(\sigma_s)], \hspace{2mm} x>x_0
\end{eqnarray}
Here the denominator measures the separation between the centers of the
spheres in units of $\sigma_s$,  and the prefactors $a_0(\eta_s)$ and $a_1(\eta_s)$ can be
calculated from the Percus-Yevick bulk pair direct correlation function \cite{ROTH00}. The amplitude
$A_p(\eta_s)$ and phase $\Theta_p(\eta_s)$ are chosen such that the
depletion potential and its first derivative are continuous at $x_0$.
They are weakly dependent on the size ratio. Fig.~\ref{fig:rothpot}
shows the form of the potential for the size ratio $q=0.1$ at a
selection of values of $\eta_s$.

The RED potential was tested in ref. \cite{ROTH00} by comparing with
computer simulation results for hard sphere mixtures and was found to
perform well for volume fractions of the small particles in the range
$0<\eta_s<0.3$. In the simulations to be described below, we use a
truncated version of the RED potential, cutoff at $x=0.3$, and with no
correction. Furthermore we shall employ the potential at finite volume
fractions $\eta_b$ of the large particles, but assume the potential
remains two-body in form, being parameterised by the reservoir volume
fraction of small particles $\tilde\eta_s$, which plays a role similar
to a chemical potential. Clearly $\tilde\eta_s\to \eta_s$ in the limit
$\eta_b\to 0$. We note that approximate expressions exist which allow
one to convert from $\tilde\eta_s$ to $\eta_s$ at finite $\eta_b$
\cite{DIJKSTRA99}, at least in the monodisperse case.

The second potential that we have studied, albeit to a lesser extent
and solely in the monodisperse context, is that due to G\"{o}tzelmann
et al \cite{GOETZELMANN98,GOETZELMANN99}. In contrast to the DFT-based
RED potential, this was derived purely within the framework of the
Derjaguin approximation, although it too is expressed as a series
expansion. We shall employ it in the truncated form studied by
Dijkstra, van Roij and Evans \cite{DIJKSTRA99}, and refer to as the DRE
potential: 

\begin{eqnarray}
\label{eq:goetzelmann}
\beta V_{\rm eff}(r_{ij}) &=& - \frac{1 + q}{2 q} \left[3 \lambda^2 \tilde\eta_s + (9 \lambda+ 12 \lambda^2) \tilde\eta_s^2\right.\\ \nonumber
                      &\:& + (36 \lambda + 30 \lambda^2) \tilde\eta_s^3 \left.\right]; \:\: -1 < \lambda <0
\end{eqnarray}
where $\lambda=x-1$. 

\begin{figure}[h]
\includegraphics[type=eps,ext=.eps,read=.eps,width=8.0cm,clip=true]{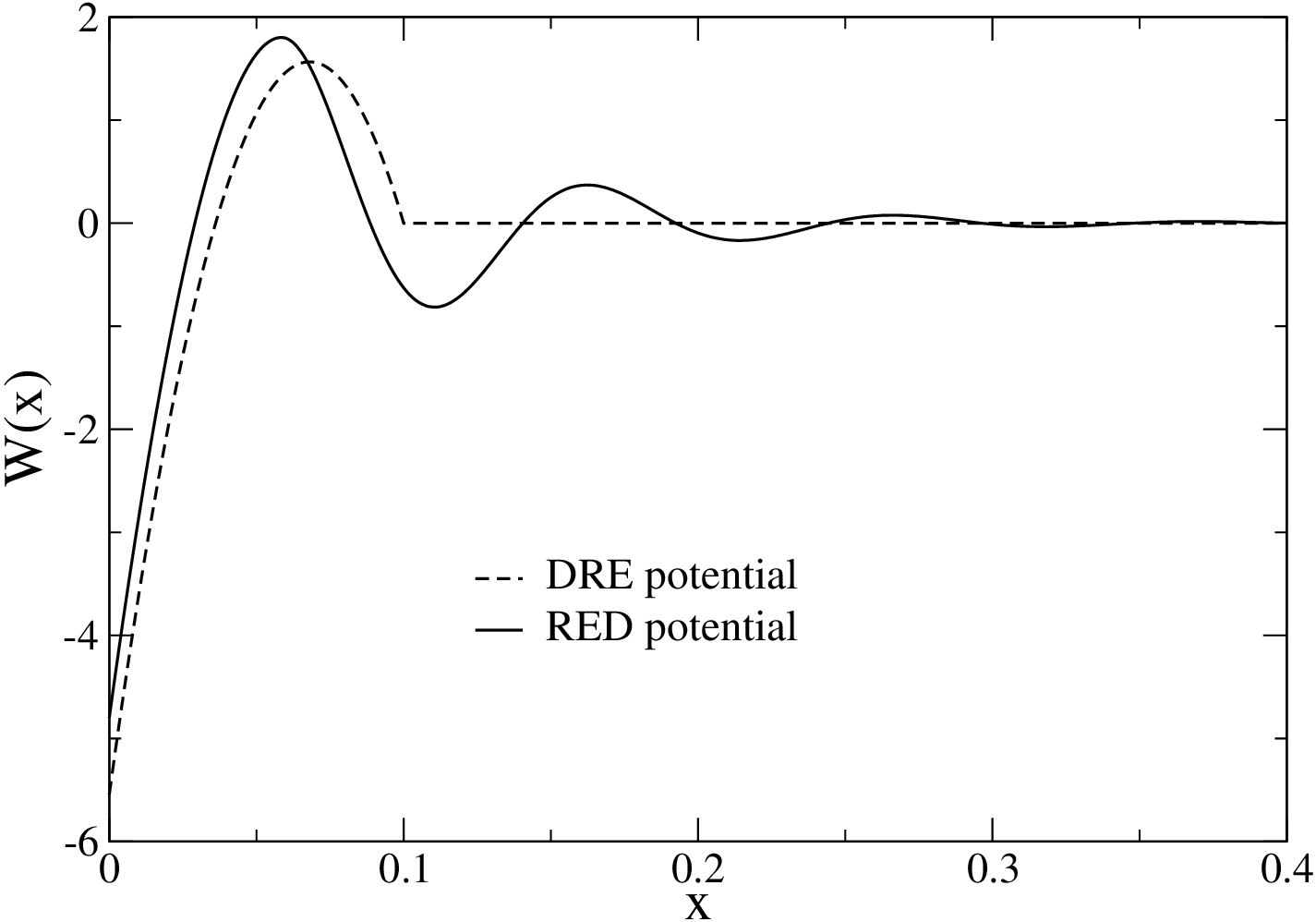}
\caption{Comparison of the RED and DRE potentials for $q=0.1$
and $\tilde\eta_s=0.3$.}
\label{fig:potcomp}
\end{figure}

Although both potentials have a qualitatively similar form at short
range, the DRE potential neglects the correct damped oscillatory decay
at larger particle separations. A comparison of the two potentials for
size ratio $q=0.1$ and $\tilde\eta_s=0.3$ is shown in fig.~\ref{fig:potcomp}.

\subsection{Incorporating polydispersity}

\label{sec:poly}

The key to incorporating polydispersity into the above framework is to
generalize the form of the depletion potential for the case of two
interacting large particles of {\em different} radii $R_1$ and $R_2$. 
This is readily achieved by appeal to the Derjaguin approximation
\cite{DERJAGUIN34,RUSSEL89}, which relates the depletion force between
two spheres of different radii ($R_1$ and $R_2$) to the potential
between two flat plates:

\begin{equation}
F_{ss}=2 \pi \frac{R_1 R_2}{R_1+R_2} U_{ww}(h)\:.
\end{equation}
The approximation also yields the force between a sphere (of radius $R$) and a wall:

\begin{equation}
F_{sw}=2 \pi R  U_{ww}(h)
\end{equation}

Clearly, if the two spheres in the first case have equal radii ($R_1=R_2$), then 

\begin{equation}
F_{ss}=\pi R U_{ww}(h), 
\end{equation}
giving the well known result $F_{sw}= 2 F_{ss}$. 

Returning to the depletion potential of eq.~\ref{eq:rothpot}, the
above considerations prompt one to write:

\begin{equation}
W=\frac{R'_1 R'_2}{R'_1+R'_2} \bar{W}
\end{equation}
where
\begin{eqnarray}
R_1^\prime&=&\frac{R_1+R_s}{R_s}\\
R_2^\prime&=&\frac{R_2+R_s}{R_s}
\end{eqnarray}
It is readily verifiable that in the limiting cases $R_1\to\infty$ and 
$R_1\to R_2$, one recovers the expression of Roth {et al} \cite{ROTH00}
(eq. ~\ref{eq:rothpot}) with $\epsilon=2$ and $\epsilon=1$
respectively.

\begin{figure}[h]
\includegraphics[type=eps,ext=.eps,read=.eps,width=8.0cm,clip=true]{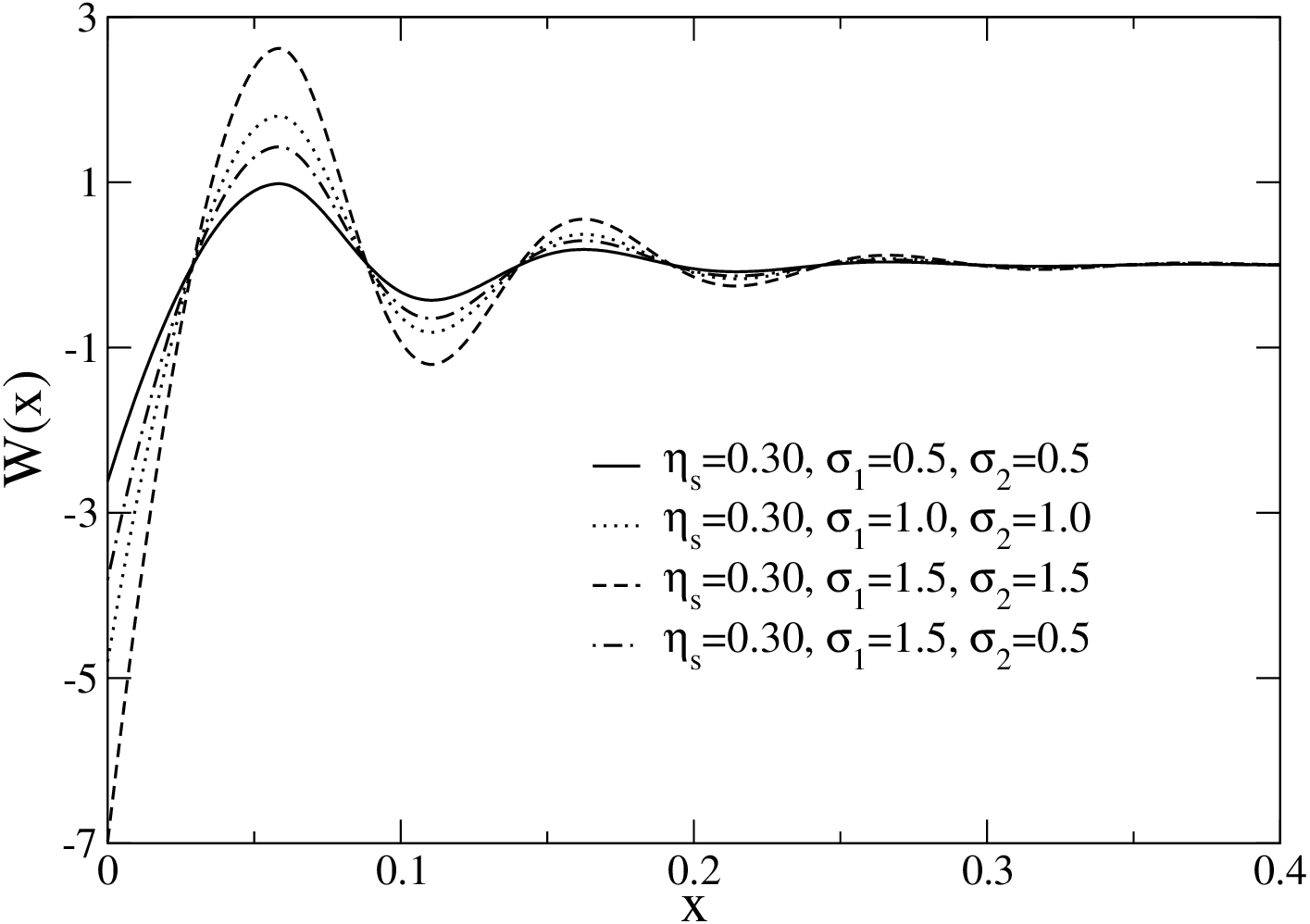}
\caption{The form of the RED depletion potential for four combinations of
the pair radii. In all cases we have set $\sigma_s=0.1$}
\label{fig:polypot}
\end{figure}

With regard to the effect of this generalization on the parameterized
form of the potential, we note firstly that the quantity $x$ remains
unaffected because it is simply the distance from contact in units of
$\sigma_s$. However, in the asymptotic part (eq.~\ref{eq:asympt}), the
denominator $\sigma_s^{-1}(\sigma_b+h)$ measuring the separation between
the centers of the spheres in $\sigma_s$ units is given in the
polydisperse case by $\sigma_s^{-1}[(\sigma_1+\sigma_2)/2+h]$. 
Additionally, while in the monodisperse limit the location of the first
maximum of the potential (at which the asymptotic behaviour is presumed
to set in) is given by $x_0=\sigma_s^{-1}(\sigma_s+\sigma_b)$, in the
polydisperse case one has instead
$x_0=\sigma_s^{-1}[\sigma_s+(\sigma_1+\sigma_2)/2]$. 
Fig.~\ref{fig:polypot} gives some examples of the influence of
dissimilar particle sizes on the depletion potential.

\begin{figure}[h]
\includegraphics[type=eps,ext=.eps,read=.eps,width=8.0cm,clip=true]{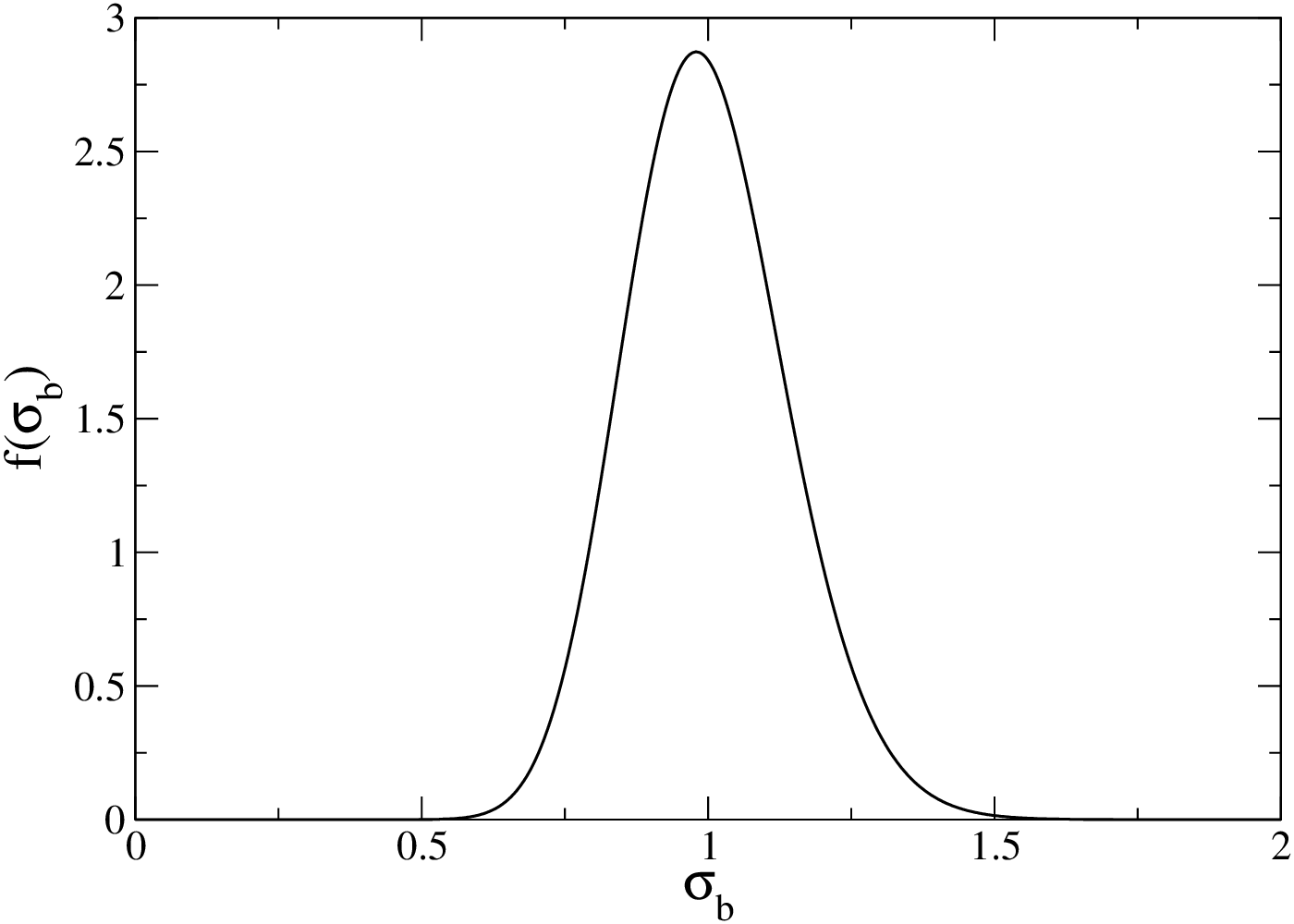}
\caption{The imposed form of $f(\sigma_b)$, corresponding to a Schulz
distribution (eq. \protect\ref{eq:schulz}) with $z=50$. The diameters $\sigma_b$
of the large particles are measured in units of $\bar\sigma_b=1$}
\label{fig:schulz}
\end{figure}

In the present work, we address the situation in which the large
particles exhibit a continuous variation of sizes.  In order to
quantify the form of the polydispersity, we label each particle by the
value of its diameter $\sigma_b$. The system can then be described in
terms of a density distribution $\rho(\sigma_b)$ measuring the number
density of particles of each $\sigma_b$. Experimentally, the
distribution of colloidal particles sizes in a system is (in general)
{\em fixed} by the synthesis of the fluid. To reflect this situation in
our simulations, we assign $\rho(\sigma_b)$ an ad-hoc prescribed functional
form, which we choose to be of the Schulz type \cite{SCHULZ39} defined
by the normalized distribution function:

\begin{equation}
f(\sigma_b)= \frac{1}{z !} \left(\frac{z+1}{\bar{\sigma_b}}\right)^{z+1} \sigma_b^z \exp{\left[-\left(\frac{z+1}{\bar{\sigma_b}}\right) \sigma_b\right]}\:.
\label{eq:schulz}
\end{equation}
Here $z$ is a parameter which controls the width of the distribution,
while $\bar\sigma_b\equiv 1$ sets the length scale. We have elected to
study the case $z=50$, for which the corresponding form of
$f(\sigma_b)$ is shown in fig.~\ref{fig:schulz}. The associated degree
of polydispersity is defined as the normalized standard deviation of
the size distribution:

\begin{equation}
\delta =\frac{ \langle \left(\sigma_b - \bar{\sigma_b}\right)^2\rangle^{\frac{1}{2}}}{\bar{\sigma_b}}
\end{equation}
For the Schulz distribution one finds $\delta=1/\sqrt{z+1}$. With
$z=50$, this formula yields $\delta\approx 14\%$. Note that for
computational convenience, lower and upper cutoffs were imposed on the
range of allowed particle sizes $\sigma_b$. These were chosen such that
$0.5 \le\sigma_b \le 1.5$. 

The imposed density distribution is related to $f(\sigma_b)$ by 

\begin{equation}
\rho(\sigma_b)=\rho_b^0f(\sigma_b)
\label{eq:dendist}
\end{equation}
where $\rho_b^0$ is the average number density of large particles.
Since $f(\sigma_b)$ is fixed, the form of $\rho(\sigma_b)$ is
parameterized solely by $\rho_b^0$, variations of which correspond (at
a given $\tilde\eta_s$) to traversing a ``dilution line'' in the full finite
dimensional phase diagram \cite{SOLLICH02}. Although this
parameterization provides the operational basis for scanning the
dilution line, we shall (in accordance with convention) quote our
results in terms of the overall {\em volume fraction} of the large
particles. The latter is related to the density distribution via

\begin{equation}
\eta_b=\int_0^\infty\frac{\pi}{6}\sigma_b^3\rho(\sigma_b)d\sigma_b\:.
\end{equation} 

Finally in this section, we note that within the polydisperse context,
the distribution of sizes of the large particles implies that the size
ratio $q$ can only be defined in terms of an average. Accordingly we
take $q\equiv\sigma_s/\bar\sigma_b$.

\section{Computational Methods}

Monte Carlo simulations were performed within the grand canonical
ensemble GCE using the methods described in
refs.~\cite{WILDING04A,WILDING04B,WILDING03,WILDING02}. Here we briefly outline
principal elements of the strategy and refer the interested reader to
those papers for a fuller description. 

Within the GCE framework, the density distribution $\rho(\sigma_b)$ is
obtained as an ensemble average over an instantaneously fluctuating
distribution. The form of $\rho(\sigma_b)$ is controlled by the
conjugate chemical potential distribution $\mu(\sigma_b)$, which was
tuned (cf ref. \cite{WILDING02}) at all points in the phase diagram
such as to yield the desired Schulz shape $f(\sigma)$
(eqs.~\ref{eq:schulz}, and scale $\rho_b^0$ \ref{eq:dendist}). This
tuning was achieved by joint use of the non-equilibrium potential
refinement (NEPR) method \cite{WILDING03}, coupled with histogram
extrapolation \cite{FERRENBERG88} in terms of $\mu(\sigma_b)$. It
should be noted, however, that the DRE and RED depletion potentials do
not lend themselves to histogram extrapolation with respect to the
model parameter $\tilde\eta_s$ which controls the form of the interaction
potential. This is because $\tilde\eta_s$ does not appear as an overall scale
factor in the Hamiltonian, a situation which contrasts, for example, to
temperature reweighting in simpler potentials such as Lennard-Jonesium.
Consequently in order to scan the phase diagram with respect to changes in
$\tilde\eta_s$, separate simulations were utilized in each instance. 

Our principal aim is a determination of the polydispersity-induced
shifts in the critical point parameters of the model depletion
potentials. To this end we have employed a crude version of the
finite-size scaling (FSS) analysis described in ref. \cite{WILDING95}.
The analysis involves scanning the range of $\rho_b^0$ and $\tilde\eta_s$
until the observed probability distribution of the fluctuating
instantaneous volume fraction of large particles $p(\eta_b)$, matches
the independently known universal fixed point form appropriate to the
Ising universality class in the FSS limit. Owing to the relatively
large depth of the interparticle potential well (see
fig.~\ref{fig:potcomp}) compared to eg. the Lennard-Jones (LJ)
potential, the acceptance rate for particle insertions and deletions
was found to be very low, resulting in extended correlation time for
the density fluctuations. Consequently we were able neither to study a
wide range of system sizes nor obtain data of sufficient statistical
quality to permit a more sophisticated FSS analysis. Nevertheless it
transpires that our estimated uncertainties on the critical point
parameters are sufficient to resolve the polydispersity-induced trends
in the critical point parameters that we set out to identify.

\section{Simulation Results}

Our results are divided into three sections. Firstly we locate the
fluid-fluid critical point for both the RED and DRE potentials in the
monodisperse limit. Moving on to the polydisperse case, we determine
the effect of the added polydispersity on the critical point
parameters. Finally we use finite-size scaling to estimate the locus of
the cluster percolation threshold in both the mono- and poly-disperse
cases.

\subsection{Monodisperse limit}

\subsubsection{Critical region}

\begin{figure}[h]
\includegraphics[type=eps,ext=.eps,read=.eps,width=8.0cm,clip=true]{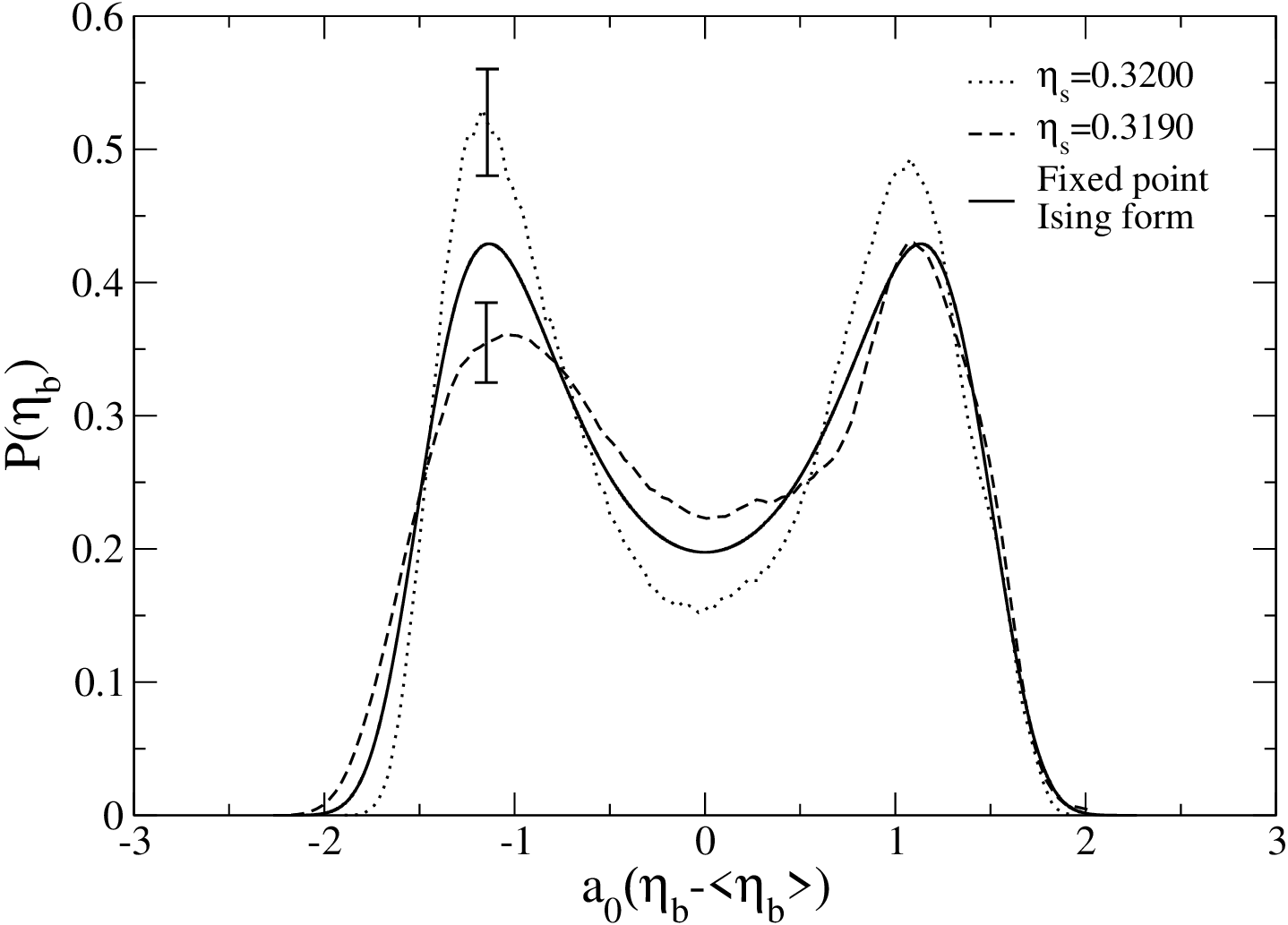}
\caption{Estimates of the order parameter distribution $p(\eta_b)$ at
$\tilde\eta_s=0.3200$ and $\tilde\eta_s=0.3190$, for $V=(5.2\bar\sigma)^3$.
Representative error bars are shown. Also included is the fixed point
Ising magnetisation distribution. All distributions are scaled to unit
norm and variance via the non-universal scale factor $a_0$.}
\label{fig:monocrit}
\end{figure}

As outlined above, we have tuned the values of $\tilde\eta_s$ and $\mu$ until
the measured probability distribution of the volume fractions of the
large particles matched (as far as possible given the computational
complexity of this problem) the universal Ising fixed point form.
Fig.~\ref{fig:monocrit} shows distributions obtained in this way for
the case of the RED potential with $q=0.1$ at $\tilde\eta_s=0.3200$ and
$\tilde\eta_s=0.3190$. Although the statistical quality is not particularly
good, comparison of the forms of the distributions with that of the 
fixed point form indicates that the given values of $\tilde\eta_s$ straddle 
criticality, permitting the estimate $\tilde\eta_s^{\rm crit}=0.3195(5)$.
This estimate for the RED potential critical point, together for that
for $q=0.05$, and the corresponding estimates for the DRE potential are
presented in tab.~\ref{tab:monocrit}.

\begin{table}
{\center RED potential}\\
\begin{tabular}{ccc}
\hline
$q$&$\tilde\eta_s^{\rm crit}$&$\eta_b^{\rm crit}$  \\
0.1  & 0.3195(5)& 0.274(10)\\
0.05 & 0.1765(5)& 0.289(15)\\
\hline
\end{tabular}

\vspace*{1cm}

DRE potential\\
\begin{tabular}{c c c}
\hline
$q$ & $\tilde\eta_s^{\rm crit}$& $\eta_b^{\rm crit}$\\
0.1 & 0.255(15)  & 0.286(15)  \\
    & [0.289] & [0.223] \\
0.05 & 0.151(1)  & 0.271(15) \\
    & [0.165] & [0.235] \\
\hline
\end{tabular}
\caption{Estimates of critical point parameters, obtained using the
methods described in the text. Values estimated from
the data of ref.\protect\cite{DIJKSTRA99} are given in square brackets.}
\label{tab:monocrit}
\end{table}

With regard to the results, of tab.~\ref{tab:monocrit}, we note that for
a given potential form, the estimates of $\eta_b^{\rm crit}$ appear
rather insensitive to the value of $q$. We further note that for a given
$q$ there is a substantial shift in $\tilde\eta_s^{\rm crit}$ between
the two forms of the depletion potential. The latter finding is perhaps
not too surprising given the significant difference in the contact value
and range of the well depth of the RED and DRE potentials, as well as
the rather radical truncation made by the DRE potential, of the long
ranged oscillatory part of the interactions (cf.
fig.~\ref{fig:potcomp}). Indeed the sensitivity of phase behaviour to
the depletion potential well depth and range has been emphasised by
Germain {\em et al} \cite{GERMAIN04}, albeit in the context of
fluid-solid coexistence.

We also note significant discrepancies between our estimates of the
critical point and those of Dijkstra et al \cite{DIJKSTRA99} for the DRE
potential. Although no error bars are quoted in ref.\cite{DIJKSTRA99},
it seem likely to us that this discrepancy is statistically significant.
Its source may be traceable to the use in \cite{DIJKSTRA99} of indirect
free energy measurements to obtain the phase diagram, in contrast to the
generally more accurate direct grand canonical FSS approach employed
here. Interestingly, our estimates for the critical $\tilde\eta_s$
values lie much closer than those of ref.~\cite{DIJKSTRA99} to the
results of a computation using integral equation theory of both the
depletion potential and its phase behaviour \cite{CLEMENT}.

\begin{figure}[h]
\includegraphics[type=eps,ext=.eps,read=.eps,width=8.0cm,clip=true]{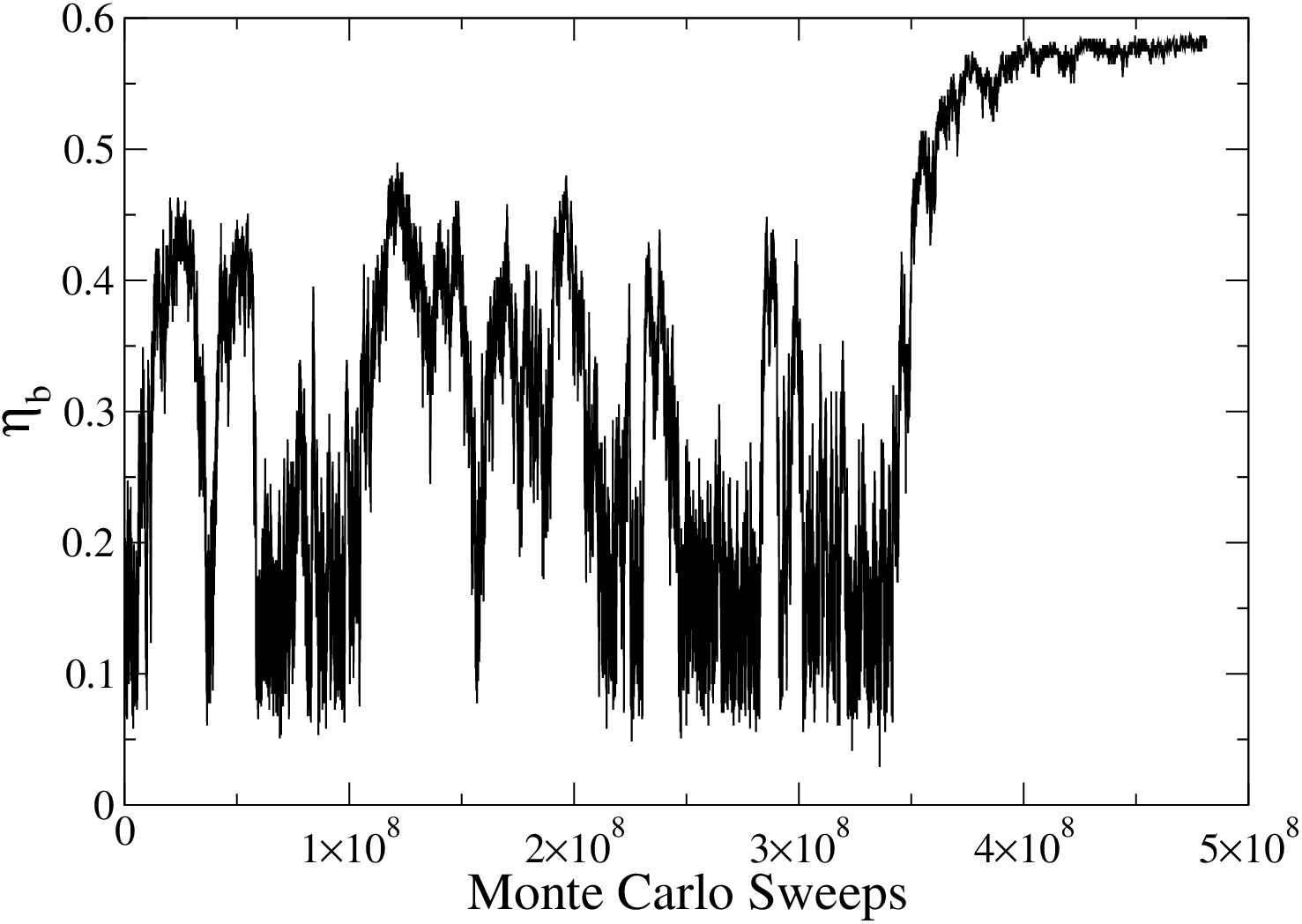}
\caption{Time evolution of the system volume fraction near the parameters
of the metastable critical point. Eventually, the system freezes spontaneously.}
\label{fig:freeze}
\end{figure}

In ref.~\cite{DIJKSTRA99} it was demonstrated via free energy
measurements that the fluid-fluid critical point for the DRE potential
is metastable with respect to freezing. While we have not attempted
to perform a systematic study of the freezing transition in the present
work, our simulations confirm the metastability in so far as some runs
were observed to freeze into an f.c.c. crystal structure.  An example
of the time evolution of the density in such a run for the DRE
potential is shown in fig.~\ref{fig:freeze}. Such freezing was also
observed for the RED potential indicating that here too the critical
point is metastable with respect to crystallization. As an aside, we
note that from a computational point of view, our observation of
freezing within the grand canonical ensemble is somewhat remarkable
since the algorithm becomes very inefficient at crystal densities. The
key factor in achieving this in the present case is the
inclusion--alongside the standard insertions, deletions and resizing
moves--of particle displacement moves. Without the latter, the system
was not observed to crystallize on simulation time scales. Another
apparent factor controlling the ease of freezing appears to be whether
the crystal lattice parameter is commensurate with the choice of system
box size. We further note that our frozen structures do not attain the
near-close packing densities observed in ref.~\cite{DIJKSTRA99}. This is
due to the presence of defects in the frozen configurations. 

Notwithstanding the eventual relaxation to a crystalline state, our
systems were usually found to remain metastable for a period of time
sufficient for us to collect useful data in the critical region.
Unfortunately, the freezing became unmanageable when we attempted to
obtain data in the fluid-fluid coexistence region. As noted by other
authors \cite{DIJKSTRA99,LOVERSO05} the coexistence curve of depletion
potentials appears to be rather flat near the critical point. Thus even a
modest excursion into the two phase region results in high liquid
densities, which in our experience froze very rapidly.

\subsection{Polydisperse case}

\label{sec:polycrit}

\begin{figure}[h]
\includegraphics[type=eps,ext=.eps,read=.eps,width=8.0cm,clip=true]{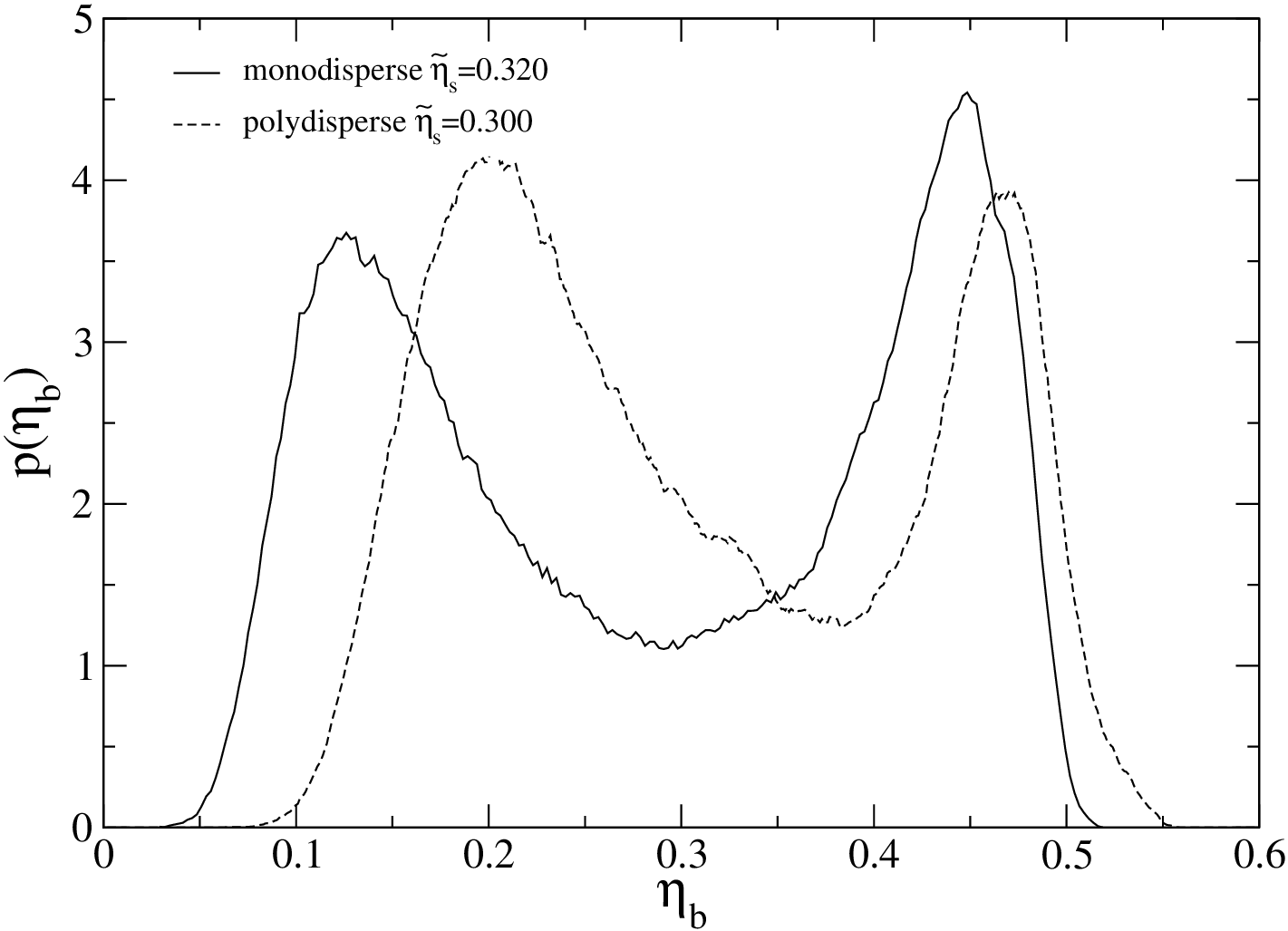}
\caption{Comparison of the distribution $p(\eta_b)$ at the estimated
critical parameters in the monodisperse and polydisperse systems. In both cases the size ratio $q=0.1$ and the system size is $V=(6\bar\sigma)^3$}
\label{fig:REDpeta}
\end{figure}

Turning now to the polydisperse case, we have obtained the critical
point parameters in a manner similar to that employed in the monodisperse
limit. Fig. ~\ref{fig:REDpeta} shows the comparison of our
estimates for the critical point distribution $p(\eta_b)$ for the RED
potential in both the mono and polydisperse cases with $q=0.1$. Clearly
the distribution for the polydisperse case is substantially shifted to
higher volume fractions compared to that for the monodisperse case.  Owing
to the slow fluctuations of $\eta_b$, the true average of
these distributions could not be determined to high precision. However,
the peak positions are rather insensitive to the fluctuations, and on
the basis of critical point universality one expects that given
sufficient statistics, the form of the distributions should become
symmetric. One can therefore estimate $\eta_b^{\rm crit}$ from the
average of the peak positions. The results of so doing are summarized
in tab.~\ref{tab:polycrit}, from which one discovers that the principal
influence of polydispersity on the critical point parameters is a
significant decrease in $\tilde\eta_s^{\rm crit}$  with respect to its
monodisperse value, and a significant increase in $\eta_b^{\rm crit}$.
For the form and degree of polydispersity we have studied,  the
decrease in $\tilde\eta_s^{\rm crit}$ is about $6\%$, while the concomitant
increase in $\eta_b^{\rm crit}$ is about $17\%$.

\begin{table}
\begin{tabular}{ccc}
\hline
&RED potential\\
$q$ & $\tilde\eta_s^{\rm crit}$ & $\eta_b^{\rm crit}$  \\
0.1  & 0.300(1)&0.336(15)\\
0.05 & 0.1655(5)&0.345(5)\\
\hline
\end{tabular}
\caption{Estimated critical point parameters for the model polydisperse system
described in sec.~\protect\ref{sec:polycrit}}
\label{tab:polycrit}
\end{table}

\begin{figure}[h]
\includegraphics[type=eps,ext=.eps,read=.eps,width=8.0cm,clip=true]{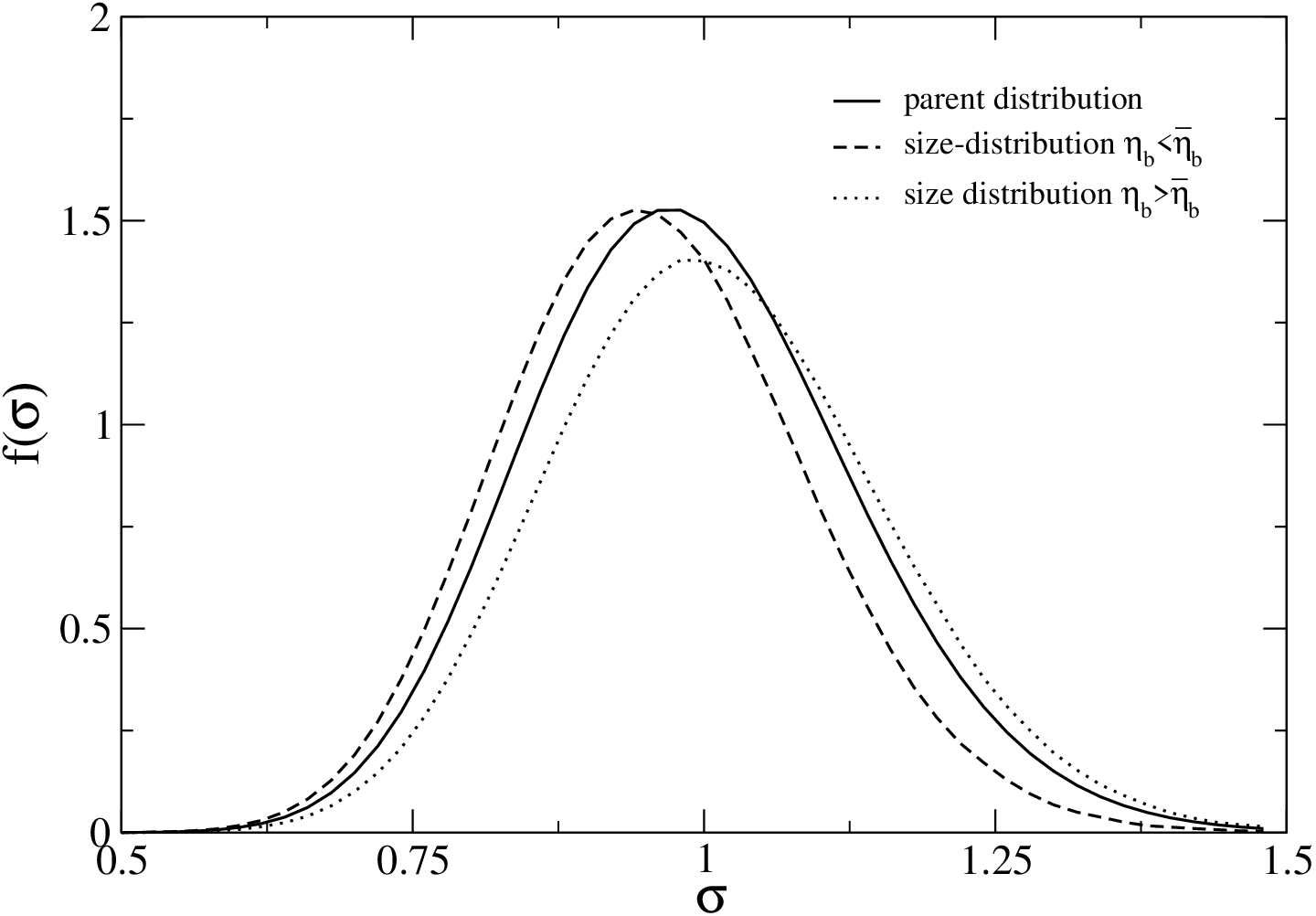}
\caption{The normalized distribution of particle sizes for
instantaneous volume fractions $\eta_b$ below (dashed line) and above
(dotted line ) the average value, close to the critical point. Also
shown (solid line) in the overall Schulz ``parent'' distribution }
\label{fig:fraction}
\end{figure}

The influence of polydispersity on the near critical point phase
behaviour is further observable in terms of particle size fractionation
effects. Specifically, when fluctuations of the instantaneous value
volume fraction $\eta_b$ exceed their average value, the distribution
of particle sizes is shifted to larger diameters; and conversely for
fluctuations of $\eta_b$ to values lower than the average. The scale of
the effect is shown in fig.~\ref{fig:fraction} for $q=0.1$ at the
estimated critical point parameters. The presence of such fractionation
implies that the critical point need not lie at the apex of the cloud
curve that marks the onset of phase separation \cite{SOLLICH02}. Indeed
we did observe some evidence for a weak separation of cloud and shadow
curves at $\tilde\eta_s=\tilde\eta_s^{\rm crit}$, although precise quantification
of the effect was complicated by a noticibly increased tendency of the
polydisperse system to relax to a high density state following a
fluctuation to high density. The nature of this relaxation closely
resembled that observed in the monodisperse case (cf.
fig.~\ref{fig:freeze}) and indeed visualization of the arrested
configurations revealed some evidence of crystalline order, albeit with
a high concentration of defects. We caution, however, that these
findings should not be interpreted as providing strong evidence for a
freezing of the polydisperse system because it was not
possible to ensure that the overall density distribution remained on
the dilution line during the relaxation to the high density state. 

\subsection{Percolation threshold}

Percolation is a necessary, though not sufficient condition for
gelation and dynamical arrest in colloidal systems. Since gelation can
affect the ability of experiments to observe equilibrium phase
behaviour in general, and specifically fluid-fluid phase separation, it is
important to determine the location of the percolation line in the
phase diagram and its relationship to the fluid-fluid critical point.
Additionally, it is of interest to ask to what extent this relationship
is affected by polydispersity.

\begin{figure}[h]
\includegraphics[type=eps,ext=.eps,read=.eps,width=8.0cm,clip=true]{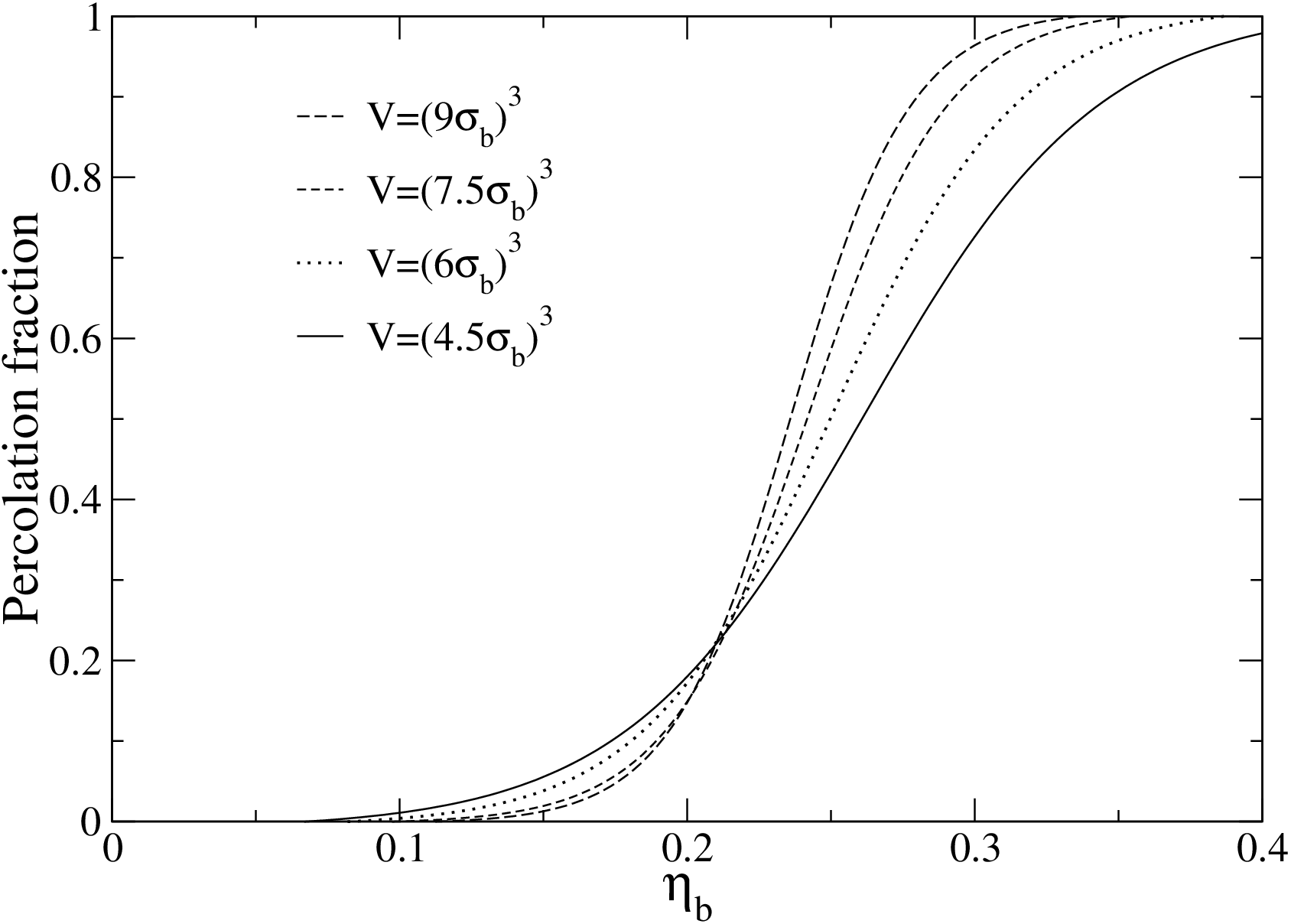}
\caption{Fraction of percolating configurations as a function of $\eta_b$ for
the RED potential in the monodisperse limit. The potential parameters
are $\tilde\eta_s=0.28$, $q=0.1$, and data are shown for four system sizes.}
\label{fig:percfss}
\end{figure}

In order to locate the percolation threshold, it is necessary to
identify pairs of particles that are `bonded' and check for spanning of
clusters of such particles. However, in contrast to lattice models or
fluid systems such as the adhesive hard sphere model, the definition of
a `bond' in systems with continuous potentials is somewhat ambiguous.
We therefore adopt a criterion which derives from that used for cluster
identification in spin models. Specifically, we determine the
interaction energy $u$ between all pairs of particles and assign a bond
with probability $p_{\rm bond}=1-\exp(\beta u)$. Clusters of bonded
particles are then identified using the algorithm of Hoshen and
Kopelman \cite{HOSHEN76}. In ref. \cite{MILLER03} Miller and Frenkel
identified the percolation threshold with those values of the model
parameters for which the proportion of configurations containing a
spanning cluster is $50\%$. However, finite-size scaling arguments
\cite{BINDER} show that better estimates may be obtained by examining
the finite-size behaviour of plots of the fraction of spanning clusters
as a function of $\eta_b$.  An example of such a plot is shown in
fig.~\ref{fig:percfss} for the RED potential in the monodisperse case
for $\tilde\eta_s=0.28$. Data are shown for $4$ system sizes, and indicate
that there is a well defined intersection point at $\eta_b\approx
0.21$. This intersection point provides a good measure of the
percolation threshold in the thermodynamic limit  \cite{BINDER}. By
contrast, application of the $50\%$ criterion to data for a single
system size can considerably overestimate the percolation threshold. 

\begin{figure}[h]
\includegraphics[type=eps,ext=.eps,read=.eps,width=8.0cm,clip=true]{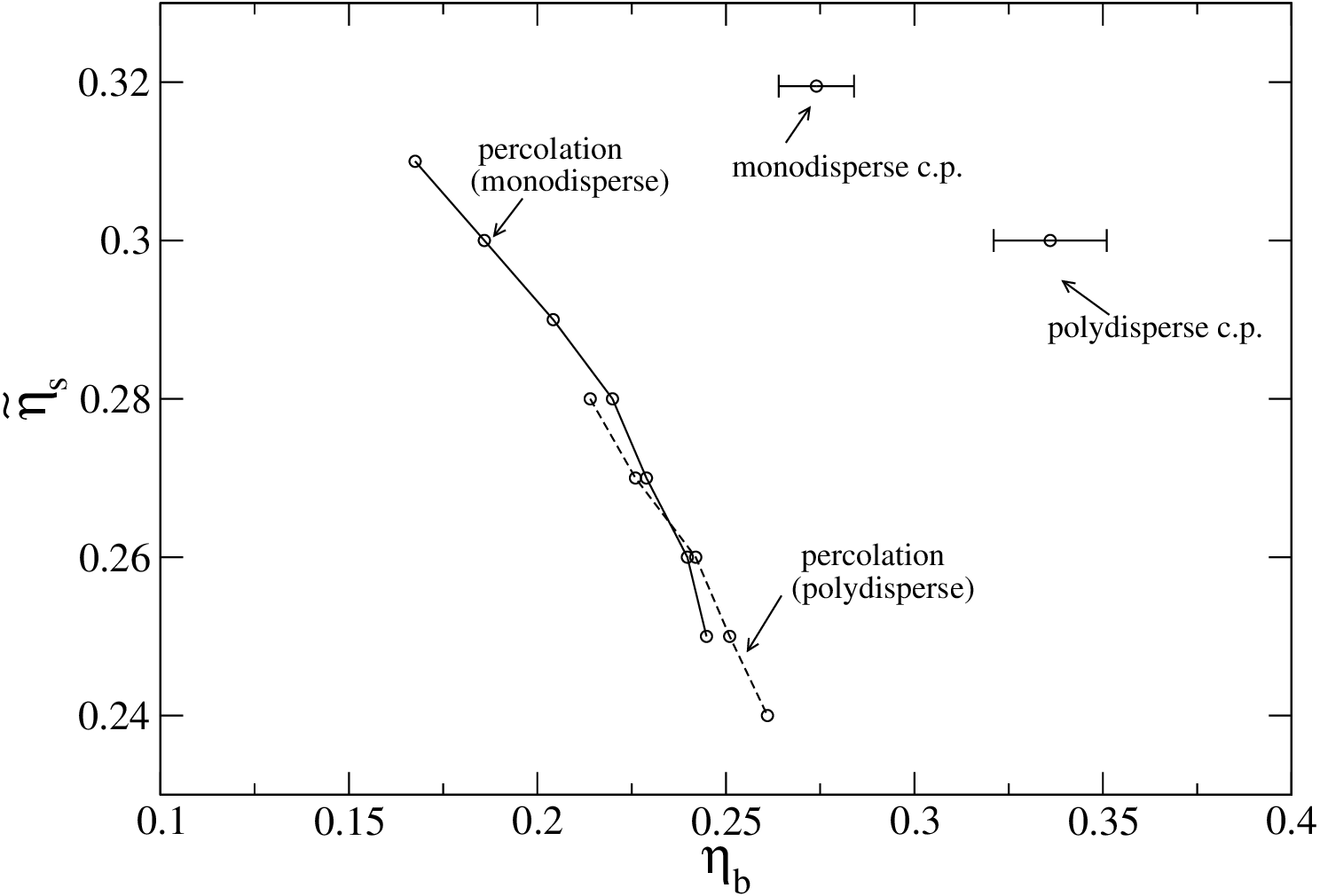}
\caption{Percolation line for the RED potential ($q=0.1$) for both the
monodisperse and polydisperse cases, as determined by the
method described in the text. Statistical errors are comparable with
the symbol sizes; lines are guides to the eye. The system size in both
cases was $L=(6\bar\sigma)^3$. Also shown are the estimated critical
point parameters (cf. sec.~\protect\ref{sec:poly}).}
\label{fig:percol}
\end{figure}

Percolation lines were determined using this intersection method for
the monodisperse and polydisperse RED potential at $q=0.1$. They are
shown in fig,~\ref{fig:percol} together with our estimates of the
critical point parameters. One sees that in both cases the critical
point lies well within the percolation regime, though much more so for
the polydisperse system than for the monodisperse system.

\section{Linking to the adhesive hard sphere model}

A simple yet general method for finding two potentials that are 
`equivalent' in a corresponding states sense, is to match their second
virial coefficient $B_2$ \cite{VLIEGENTHART00,NORO00,MILLER03}:

\begin{equation}
B_2=-2 \pi \int^{\infty}_{0} \left(e^{-\beta u(r)}-1  \right) r^2 dr\:.
\end{equation}

Here we compare the value of $B_2$ for the RED and DRE depletion
potentials in the monodisperse limit, with that of the adhesive hard
sphere (AHS) model \cite{BAXTER68}. The latter comprises hard particles
which experience a finite attraction only at contact, the strength of which
is controlled via a `stickiness' parameter $\tau$. The  overall
interaction can be written 

\begin{equation}
e^{-\beta u(r)}=\Theta (r-\sigma)+\frac{\sigma}{12
\tau}\delta(r-\sigma)\:,
\end{equation}
with $r$ the separation of particle centers and $\sigma$ the particle
diameter. The second virial coefficient follows as 

\begin{equation}
B_2^{AHS}=\frac{2\pi}{3} \sigma^3  \left(1-\frac{1}{4 \tau}\right)\:.
\label{eq:B2AHS}
\end{equation}

The AHS model exhibits a fluid-fluid phase transition, the critical
point of which has been estimated to occur \cite{MILLER04b} at 
$\tau_c=0.1133(5), \rho_c=0.508(10)$. This value of $\tau_c$ implies
that for the AHS model at criticality, $B_2^{\rm crit}=-4.826 v_0$ with
$v_0=(4/3)\pi\sigma^3$. It is therefore of interest to assess whether,
via the matching of $B_2$ values for the RED and DRE potentials to
that of the critical AHS model, reasonable predictions can be made for
the critical point parameters of the depletion potentials. To this end
we have numerically evaluated $B_2$ across a range of $\tilde\eta_s$ values
for each depletion potential and $q$ value of interest. By so doing
we could determine that value of $\tilde\eta_s$ for which $B_2$ matched the value
$B_2^{\rm crit}=-4.826 v_0$. Table.~\ref{tab:ahscomp} shows the
resulting predictions for $\tilde\eta_s^{\rm crit}$ for two values of $q$,
together with our simulation estimates. Clearly, in each instance, the
agreement is remarkable.

\begin{table}
\begin{tabular}{ccc}
\hline
&Monodisperse RED potential&\\
q&$\tilde\eta_s$ predicted & $\tilde\eta_s$ simulation \\
0.1& 0.320&0.3195(5)\\
0.05&0.177&0.1765(5)\\
\hline
&Monodisperse DRE potential&\\
q&$\tilde\eta_s$ predicted & $\tilde\eta_s$ simulation \\
0.1& 0.256 & 0.255(15)\\
0.05&0.151 & 0.151(1)\\
\hline
&Polydisperse RED potential&\\
q&$\tilde\eta_s$ predicted & $\tilde\eta_s$ simulation \\
0.1& 0.310 &0.300(1)\\
0.05&0.172 &0.1655(5)\\
\hline
\end{tabular}
\caption{Comparison of the simulation estimates of the critical point
parameters with the predictions arising by matching the second virial
coefficient to that of the critical AHS model.}
\label{tab:ahscomp}
\end{table}

One can attempt to extend the above approach to the polydisperse
depletion potentials. To do so, we first obtain the contribution to the
second virial coefficient for interactions between all pairs of species
$\sigma_i,\sigma_j$. The overall coefficient for the mixture can then
be  approximated as a weighted average of pairs \cite{HALL93}, where
the weight factor is the probability of interaction between a pair of
particles of size $i$ and size $j$. In our case, this is given by the
product of the corresponding values of the normalized Schulz
distribution (eq.\ref{eq:schulz}):

\begin{equation}
B_2=\int_0^\infty\int_0^\infty  f(\sigma_i) f(\sigma_j) B_2(\sigma_i,\sigma_j)d\sigma_id\sigma_j
\label{eq:polyB2}
\end{equation}

Matching to $B_2^{\rm crit}=-4.826v_0$ as before, one obtains for the
two $q$ values studied, the predictions for $\tilde\eta_s^{\rm crit}$ shown
in  tab~\ref{tab:ahscomp}. Here the agreement with the simulation
estimates is less impressive than in the monodisperse case.  Although
the absolute value of the prediction still agrees to within about $3\%$
with the simulation estimate, and the sign of the
polydispersity-induced shift in $\tilde\eta_s^{\rm crit}$ is correctly
predicted, its magnitude is underestimated by a factor of two. 

The larger relative discrepancy between the predicted and measured
$\tilde\eta_s^{\rm crit}$ may point to a breakdown in the presence of
polydispersity of the assumed model invariance of the critical $B_2$
value. Indeed one might expect such a failure because the value of
$B_2$ is based solely on the pair potential and takes no account of the
ability of a polydisperse fluid to exploit local size segregation in
order to pack more effectively than a corresponding monodisperse one.
In order to address this issue directly, one would require estimates of
critical point $B_2$ values for a polydisperse version of the AHS
model. To our knowledge no simulation estimates of the liquid-gas
transition currently exist for a polydisperse AHS model. Indeed, the
matter is complicated by the fact that there is no {\em unique} model
for polydispersity in such a system. Very recently, however, a number
of physically reasonable models for polydispersity in AHS system have
been proposed by Fantoni et al \cite{FANTONI}, who investigated the
corresponding phase behaviour using integral equation theory. From
ref.~\cite{FANTONI}, one can deduce that the presence of polydispersity
significantly {\em decreases} the magnitude of $B_2$ at the critical
point compared to the monodisperse limit. This trend in $B_2$ is of the
correct sign and overall magnitude to push the predictions for
$\tilde\eta_s^{\rm crit}$ for our depletion potentials closer to the
simulation estimates. Unfortunately since no data were reported for
exactly the same degree of polydispersity ($\delta=14\%$) studied in
the present work, no direct comparison of $B_2$ values is possible. 

In an attempt to throw additional light (albeit indirectly) on the
discrepancy between the measured and predicted critical point
parameters, we have studied the effect of introducing polydispersity on
the critical point $B_2$ value for the Lennard-Jones fluid, which is a
computationally more tractable system than the AHS model
\cite{MILLER04}. The corresponding potential is

\begin{equation}
u_{ij}=\epsilon_{ij}\left[\left(\frac{\sigma_{ij}}{r_{ij}}\right)^{12}-\left(\frac{\sigma_{ij}}{r_{ij}}\right)^{6}\right]
\end{equation}
with $\epsilon_{ij}=\sigma_i\sigma_j\epsilon$,
$\sigma_{ij}=(\sigma_i+\sigma_j)/2$ and $r_{ij}=|{\bf r}_i-{\bf r}_j|$.
The potential was cutoff for $r_{ij}>2.5\sigma_{ij}$ and no tail
corrections were applied. For the monodisperse limit, the critical
temperature occurs at $T_c=1.1876(3)$ \cite{WILDING95} and one finds
$B_2=-6.621v_0$, which lies within the range of `typical'
critical point $B_2$ values found in surveys of a wide range of model
potentials \cite{VLIEGENTHART00,NORO00}. If, on the other hand,
$\sigma$ is distributed according to a Schulz form (eq.\ref{eq:schulz})
with $z=50$, as used elsewhere in this work, simulations yield a
critical temperature of $T_c=1.384(1)$, for which (using
eq.~\ref{eq:polyB2}), one finds $B_2=-5.759v_0$, which is significantly 
smaller in magnitude that the monodisperse value.

If one now makes the (not unreasonable) assumption that given the same
form and degree of polydispersity, a comparable fractional change of
$B_2$ will ensue in other interaction potentials, one can estimate the
expected critical point value of $B_2$ for the polydisperse AHS model
(and hence also the polydisperse depletion potentials), as
$B_2=-4.826\times 5.759/6.621\approx-4.2v_0$. The resulting predictions
for the critical reservoir volume fraction of the small particles are
$\tilde\eta_s^{\rm crit}=0.304$, for $q=0.1$, and $\tilde\eta_s^{\rm crit}=0.169$
for $q=0.05$, both of which agree within error with our simulation
results. Of course in view of our assumptions, this accord may be
fortuitous, but it is nevertheless suggestive that a more detailed
assessment of the effect of polydispersity on critical point $B_2$
values in other systems (specifically the AHS model) would be a
worthwhile avenue for further study.

\section{Conclusions and discussion}

To summarize, we have determined the effect of introducing
polydispersity on the fluid-fluid critical point parameters of a model
depletion potential for highly asymmetric additive hard sphere mixtures.
For the particular realization of the polydispersity considered, the
critical point is found to shift (with respect to the monodisperse
limit) to smaller values of the reservoir volume fraction of the small
particles $\tilde\eta_s$ and to larger values of the volume fraction of
the large particle $\eta_b$. It seems reasonable to assume that the
direction of the shifts should be a general trend, common to other
functional forms and degree of polydispersity. Indeed the same trend has
also recently been observed in a study of colloidal polydispersity in
the AO model \cite{FASOLO04}. 

Beyond this, our results show that inclusion of polydispersity pushes
the whole fluid-fluid binodal deeper into the percolating regime. Since
colloidal fluids are known to form a gel \cite{BERGENHOLTZ00,CATES04}
for sufficiently high $\tilde\eta_s$, it would seem that the presence of
polydispersity increases the likelihood that direct observations of
fluid-fluid phase coexistence is complicated by dynamical arrest. 

Additionally we demonstrated that excellent predictions for the value
of $\tilde\eta_s^{\rm crit}$ follow from matching the second virial
coefficient $B_2$ of depletion potentials to the critical $B_2$ value
of the adhesive hard sphere model. The quantitative accuracy of the
predictions is undoubtedly due in large part to the very short ranged
nature of the depletion potentials; similar studies comparing critical
point $B_2$ values for a range of other potentials
\cite{VLIEGENTHART00,NORO00} did not find such a high degree of
accuracy. Nevertheless our observation should prove generally useful in
reducing the effort required to locate criticality in depletion
potentials. It is intriguing however, that the accuracy of the
predictions was reduced on incorporating polydispersity, suggesting
that (perhaps due to changes in packing ability due to local size
segregation effects), the inclusion of polydispersity in a model does
not leave $B_2$ invariant at the critical point. Comparisons of the
critical $B_2$ value for a monodisperse and polydisperse LJ fluid
confirmed a significant difference in this regard. Moreover, the
magnitude of the effect was sufficient to account for the discrepancy
in the observed and predicted $\tilde\eta_s^{\rm crit}$ for the polydisperse
depletion potential. Clearly, however, further work is called for in
order to elucidate this matter more fully.

Obviously knowledge of the shift in the critical point parameters is in
itself insufficient to determine whether polydispersity renders the
fluid-fluid transition stable with respect to fluid-solid coexistence.
Although we did observe a spontaneous relaxation of the near critical
polydisperse system to a high density state showing some crystalline
order, this finding should be treated with caution because the system
departs from the dilution line during the formation of the new state.
It would thus be interesting in future work to try to study explicitly
the effects of polydispersity on the freezing transition. As well as
providing assessment of the overall stability of the fluid-fluid
critical point, freezing in polydisperse fluids is a matter of
considerable interest in its own right. Indeed recent theoretical
calculations for the AO model indicate an increasing richness of
fluid-solid and solid-solid phase behaviour as the degree of
polydispersity is increased \cite{FASOLO04}. To tackle this
computationally, however, is a considerable challenge, but one which 
might be met by extending to polydisperse system novel computational
methods which have hitherto only be deployed in the monodisperse
context \cite{WILDING00,GRAHAM}.

Finally, we remark that while the present work has considered solely the
case of polydispersity of the large particles, the converse situation of
small particles polydispersity is clearly of interest and practical
relevance too. While there have been several theoretical studies which
have considered this scenario (see eg.
refs.~\cite{GOULDING01,SOLLICH05}), we know of no simulation studies to
date. An extension of the methods utilised here to address this case 
would doubtless be a worthwhile endeavor--one which we hope to undertake
in future work.

\acknowledgments

The authors are grateful to  R. Evans and A. Louis for helpful
discussions and advice, and to M. Dijkstra for supplying data from
ref.~\cite{DIJKSTRA99}. This work was supported by EPSRC, grant number
GR/S59208/01. JL acknowledge support from Spanish Ministerio de
Educacion, Cultura y Deporte.

\end{document}